\documentstyle[a4,psfig,amsmath,amsxtra,amsthm,amssymb,citesort]{article}

\begin{document}
\centerline{{\large\bf Spatio-temporal organization of vehicles in a}}
\centerline{{\large\bf cellular automata model of traffic with 
'slow-to-start' rule}}

\vspace{1cm}
\begin{center}
Debashish Chowdhury$^{*,\dagger}$, Ludger Santen$^{\dagger}$, 
Andreas Schadschneider$^{\dagger}$,\\
Shishir Sinha$^{*}$\footnote{Present address: Department of Physics, 
Harvard University, Cambridge, MA 02138, U.S.A.}, 
and Abhay Pasupathy$^{*}$\footnote{Present address: LASSP, Cornell 
University, Ithaca, NY 14853, U.S.A.}
\end{center}

\vspace{1cm}

\centerline{$^*$Physics Department, Indian Institute of Technology, 
Kanpur 208016, India}
\centerline{$^{\dagger}$Institute for Theoretical Physics, University of 
Cologne, D-50923 K\"oln, 
Germany} 

\vspace{1cm}

\noindent{\bf Abstract:}
The spatio-temporal organizations of vehicular traffic in 
cellular-automata models with "slow-to-start" rules  are 
qualitatively different from those in the Nagel-Schreckenberg 
(NaSch) model of highway traffic. Here we study 
the effects of such a slow-to-start rule, introduced by Benjamin, 
Johnson and Hui (BJH), on the the distributions of the 
distance-headways, time-headways, jam sizes and sizes of the 
gaps between successive jams by a combination of approximate 
analytical calculations and extensive computer simulations. We 
compare these results for the BJH model with the corresponding 
results for the NaSch model and  interpret the qualitative 
differences in the nature of the spatio-temporal organizations 
of traffic in these two models in terms of a phase separation 
of the traffic caused by the slow-to-start rule in the BJH model. 

\vspace{2cm}


\newpage

\section{Introduction}

The continuum theories of the flow of vehicular traffic
\cite{Lighthill55,Kerner97,Choi95} are  
analogous to the hydrodynamic description of fluid flow. The kinetic 
theories of vehicular traffic \cite{Prigogine71,Helbing95,Nagatani97} are
extensions of the kinetic  
theory of gases whereas the "car-following models"
\cite{HermanGazis,Bando95,Yukawa96,Nagatani96} and the 
recent particle-hopping models
\cite{Nagel92,Nagel93,Schreckenberg95,Schadschneider97a,Schadschneider98,Nagel96,Chowdhury97,Ghosh98,Chowdhury98,Schadschneider98a,Ktitarev97,Csahok94,Nagel95,Boccara97,Benjamin96,Takayasu93,Barlovic98,Schadschneider97,Wolfneu}
are analogues of the models 
of driven system of interacting particles
\cite{Schmittmann95}. Therefore, the  
theoretical techniques of fluid dynamics and non-equilibrium 
statistical mechanics have turned out to be powerful tools in the 
study of a wide variety of problems in the science and engineering 
of vehicular traffic \cite{May90,Wolf96,Schreckenberg98}. One of the
most popular "particle-hopping  
models" of vehicular traffic is the Nagel-Schreckenberg (NaSch) model 
\cite{Nagel92}, which may be regarded as a stochastic cellular automaton (CA) 
\cite{Wolfram86}, where the time-evolution of the vehicles is formulated in 
terms of a set of "rules". Several extensions of these rules have 
been suggested in recent years (see, e.g.,
\cite{Ktitarev97,Csahok94,Nagel95,Boccara97,Benjamin96,Takayasu93,Barlovic98,Schadschneider97,Wolfneu})
to make the model more realistic; in this paper we study some of the
statistical properties of one such extension \cite{Benjamin96}.  

The flux versus density of the vehicles is known as the {\it 
fundamental diagram} \cite{May90}. The {\it distance-headway} ({\bf DH}) 
is defined as the distance from a selected point on the lead vehicle 
({\bf LV}) to the same point on the following vehicle ({\bf FV}); 
usually the front edges or bumpers are selected \cite{May90}. The {\it 
time-headway} ({\bf TH}) is defined as the time interval between 
the departures (or arrivals) of two successive vehicles recorded 
by a detector placed at a fixed position on the highway \cite{May90}. The 
fundamental diagram, the DH and TH distributions, the distributions 
of the sizes of the traffic jams as well as the distribution of the 
gaps between successive jams in the NaSch model on idealized single-lane 
highways have been calculated
\cite{Nagel92,Nagel93,Schreckenberg95,Schadschneider97a,Schadschneider98,Nagel96,Chowdhury97,Ghosh98,Chowdhury98,Schadschneider98a}.
Very recently, the update 
"rules" of the NaSch model have been extended by adding a step where  
the so-called "slow-to-start" rules (which will be explained in the 
next section) \cite{Benjamin96,Takayasu93,Barlovic98,Schadschneider97}
are implemented. The fundamental diagram of  
some of these CA models with "slow-to-start" rules have been   
investigated \cite{Schadschneider97}. The main aim of this paper is to calculate, by 
a combination of analytical and numerical methods, the 
distributions of DH, TH, jam sizes and distances between jams in the 
CA model with the "slow-to-start" rule suggested by Benjamin, Johnson 
and Hui ({\bf BJH}) \cite{Benjamin96} to study the effects of the slow-to-start 
rule on the spatio-temporal organization of traffic. We interpret 
the novel features of these distributions in terms of a phase 
separation of the traffic in the BJH model. 

The NaSch model \cite{Nagel92} and the BJH model \cite{Benjamin96}, which was formulated by 
BJH by incorporating one specific type of "slow-to-start" rule in 
the updating scheme of the NaSch model, are stated in section 2 for 
the sake of completeness. The theoretical techniques, which will be 
used in our analytical calculations, and the numerical methods, 
which will be followed in our computer simulations, are also explained 
briefly in the same section. The probability distributions, namely the 
distributions of DH, TH, jam sizes and gaps between jams, with which 
we are mainly concerned in this paper, are defined in section 2. 
Our results for the DH distributions are presented in section 3. Our 
analytical as well as numerical results for the distributions of jam 
sizes and jam gaps are presented in sections 4 and 5, respectively, 
while those for the TH 
distributions are given in section 6. We make a quantitative estimation 
of the regime of validity of our analytical expressions by computing 
an appropriate correlation function in section 7. A summary of our 
results and conclusions drawn from these are given in section 8. 
Three appendices contain details of our analytical calculations.

\section{Models and theoretical techniques} 

In the particle-hopping models a lane is represented by a 
one-dimensional lattice of $L$ sites. Each of the lattice sites can 
be either empty or occupied by at most one "vehicle". If periodic 
boundary condition is imposed, the density $c$ of the vehicles is 
$N/L$ where $N (\leq L)$ is the total number of vehicles. Throughout 
this paper we shall follow the convention that the vehicles move from 
left to right (i.e., along the positive $X$-axis), so that in a given 
configuration $(a,b)$, of a pair of sites, $b$ refers to the LV and 
$a$ refers to the FV. 

\subsection{The models} 

\subsubsection{The NaSch model} 

For the sake of completeness we briefly recall the definition of 
the NaSch model \cite{Nagel92}. In this model, the speed $V$ of each vehicle 
can take one of the $V_{max}+1$ allowed {\it integer} values 
$V=0,1,...,V_{max}$. Suppose, $X_n$ and $V_n$ denote the position 
and speed, respectively, of the $n$-th vehicle. Then, $d_n = X_{n+1}-X_n$, 
is the gap in between the $n$-th vehicle and the vehicle in front of 
it at time $t$. At each {\it discrete time} step $t \rightarrow t+1$, 
the arrangement of $N$ vehicles is updated {\it in parallel} 
according to the following "rules":

\noindent {\it Step 1: Acceleration.} $V_n \rightarrow min(V_n+1,V_{max})$.

\noindent{\it Step 2: Deceleration (due to other vehicles).} 
$V_n \rightarrow min(V_n,d_n-1)$.

\noindent{\it Step 3: Randomization.} $V_n \rightarrow max(V_n-1,0)$ 
with probability $p$. 

\noindent{\it Step 4: Vehicle movement.} 
$X_n \rightarrow  X_n + V_n$. 

Non-vanishing braking probability $p$ 
is essential for a realistic modeling of traffic flow
\cite{Schreckenberg95} and, therefore,  
the NaSch model may be regarded as stochastic cellular automata
\cite{Schreckenberg98}. 

\subsubsection{\bf The BJH model} 

To our knowledge, so far three different versions of the "slow-to-start" 
rule have been formulated
\cite{Benjamin96,Takayasu93,Barlovic98,Schadschneider97} ; in this
paper we shall consider  
only the slow-to-start rule introduced by Benjamin et al. (BJH)
\cite{Benjamin96}.  
BJH modified the updating rules of the NaSch model by introducing an extra 
step where their "slow-to-start" rule is implemented. According this 
"slow-to-start" rule, the vehicles which had to brake due to the next 
vehicle ahead will move on the next opportunity only with probability 
$1-p_s$. The steps of the update rules can be stated as follows:\\

\noindent {\it Step 1: Acceleration.} $V_n \rightarrow min(V_n+1,V_{max})$.

\noindent{\it Step 2: Slow-to-start rule}: 
If $flag = 1$, then $V_n \rightarrow 0$ with probability $p_s$. 

\noindent{\it Step 3: Blockage (due to other vehicles).} 
$V_n \rightarrow min(V_n,d_n-1)$ and, then,\\ ~~ $flag = 1$ if $V_n = 0$, 
else $flag = 0$.

\noindent{\it Step 4: Randomization.} $V_n \rightarrow max(V_n-1,0)$ 
with probability $p$. 

\noindent{\it Step 5: Vehicle movement.} 
$X_n \rightarrow  X_n + V_n$.\\ 
Here $flag$ is a label distinguishing vehicles which have to obey the 
slow-to-start rule ($flag = 1$) from those which do not have to 
($flag = 0$).

For $p_s=0$ the above rules reduce to those of the NaSch model.
Note that, in both these models, if $V_{max} = 1$, a vehicle can come 
to an instantaneous halt, even at vanishingly low densities, because 
of random braking.  In contrast, for all $V_{max}>1$, vehicles do not 
stop spontaneously at sufficiently low densities for which interactions 
among the vehicles is negligibly small. Therefore, effectively free 
flow of traffic takes place when the density of vehicles is sufficiently 
low whereas high density leads to congestion and traffic jams. The 
density, $c_o$, corresponding to the maximum flux is usually called the 
{\it optimum} density. At sufficiently high densities, the BJH model 
with $V_{max} > 1$ is known to exhibit phase separation where one of 
the two phases consists of a macroscopic jam while the other corresponds 
to free flow of traffic. Moreover, the  particle-hole symmetry of the 
NaSch model for $V_{max} = 1$ is lost in the corresponding BJH model 
because of the slow-to-start rule \cite{Schadschneider97}. 

\subsection{Theoretical techniques}

We compute the DH distribution in the BJH model with $V_{max} = 1$ 
using some analytical results derived earlier \cite{Schadschneider97} within the framework 
of a car-oriented mean-field ({\bf COMF}) theory
\cite{Schadschneider97a}. Our analytical   
calculations of the distributions of jam sizes in the BJH model with 
$V_{max} = 1$ are also carried out within the framework of the same 
COMF theory. We derive analytical expressions for the distributions 
of the gaps between successive jams in the BJH model for $V_{max} = 1$ 
using a 2-cluster site-oriented mean-field ({\bf SOMF}) theory which 
leads to the exact results for the NaSch model with $V_{max} = 1$ in 
the limit $p_s \rightarrow 0$, although the results for the BJH model 
($p_s \neq 0$) are not exact. We also present an alternative derivation 
of the distributions of the gaps between successive jams in terms of a 
hybrid approach which will be explained in section 5. Finally, we 
derive the analytical expression for the TH distributions in the BJH 
model with $V_{max} = 1$ following an extension of the approach used  
earlier \cite{Chowdhury98} for the corresponding calculation for NaSch model.  

In our computer simulations we begin with a random initial
configuration and let a system of $L=1000$ sites evolve for $10^4$
steps which is long enough to ensure that the system reaches the
steady state; then we let the system evolve for $10^5$ further steps
during which we compute the properties of our interest. Then the data
are averaged over $100$ samples (i.e., $100$ initial configurations)
for each of the sets of values of the parameters.  We have analyzed
data for a wide range of values of the parameters $p$ and $p_s$.
However, we present here data mainly for the regime $p \ll p_s$ where
effects of the slow-to-start rule are the strongest.

\subsection{The distributions of our interest} 

In this section we define the distributions of our interest. 
The number of empty lattice sites in front of a vehicle is taken to be 
a measure of the corresponding DH. Suppose ${\cal P}_{dh}(k)$ denotes 
the probability of a DH $k$; more precisely, ${\cal P}_{dh}(k)$ is the 
conditional probability of finding a string of $k$ empty sites in front 
of a site which is given to be occupied by a vehicle. Note that 
${\cal P}_{dh}(0)$ is also the total fraction of vehicles which are 
simultaneously in the jammed state.

A jam of length $k$ is defined as a string of $k$ successive stopped 
vehicles, i.e.\ we are considering only compact jams. 
Similarly, when there are $k$ lattice sites between two successive 
jams, each occupied by a moving vehicle or is vacant then we say that 
there is a gap of length $k$ between the two successive jams; we are 
interested in the distribution ${\cal P}_{jg}(k)$. 

We define the TH as follows \cite{Ghosh98}. A detector is placed at one of the 
lattice sites. In order to measure the TH between a given pair of 
vehicles the detector is set to $t = 0$ instantaneously when the LV 
of the pair leaves the detector site. From then onwards the detector 
counts the number of discrete time steps up to $t = \tau$ when the FV 
leaves the same detector site; $\tau$ is the TH between the pair of 
vehicles under consideration. As soon as the FV leaves the detector 
site the detector instantaneously resets again to $t = 0$ in order to 
measure the TH between this vehicle and the vehicle following it. We 
denote the TH distribution by the symbol ${\cal P}_{th}(\tau)$.

\section{The DH distribution} 

Suppose, ${\cal P}(k)$ denotes the probability to find exactly $k$ 
empty cells in front of a vehicle. Only vehicles which have exactly
one empty cell in front might be affected by the slow-to-start rule
\cite{Schadschneider97}.
We denote the density of these vehicles by ${\tilde {\cal P}}(1)$.
The normalization requires 
\begin{equation}
\sum_{k \geq 0} {\cal P}(k) + {\tilde{\cal P}}(1) = 1.  
\label{eq1}
\end{equation}
Note that 
${\cal P}_{dh}(1) = {\cal P}(1) + {\tilde{\cal P}}(1)$, and 
${\cal P}_{dh}(k) = {\cal P}(k)$ for all $k \neq 1$. 
The symbol $g(t)$ ($\bar{g}(t)$) is the probability that the vehicle 
moves (does not move) in the next time step; therefore, 
\begin{equation}
g = q \sum_{k \geq 1} {\cal P}(k) + q q_s {\tilde {\cal P}}(1) 
\label{eq2}
\end{equation}
where $q = 1-p$ and $q_s = 1 - p_s$. Making the ansatz 
\begin{equation}
{\cal P}(k) = {\cal N} z^k  \quad (k \geq 2)
\label{eq3}
\end{equation} 
one gets the following results \cite{Schadschneider97}: 
\begin{equation}
{\cal P}(1) = \frac{z{\cal P}(0)}{p} - q_s{\tilde{\cal P}}(1) 
\label{eq4}
\end{equation}
\begin{equation}
{\tilde{\cal P}}(1) = \frac{q{\cal P}(0)(1-{\cal P}(0))}{1+p_sq{\cal P}(0)} 
\label{eq5}
\end{equation}
where 
\begin{equation}
{\cal N} = \frac{1-z}{z^2}\cdot [1 - {\cal P}(0) - {\cal P}(1) - 
{\tilde {\cal P}}(1)],
\label{eq6}
\end{equation}
\begin{equation}
z = \frac{pg}{q(1-g)}, 
\label{eq7}
\end{equation}
\begin{equation}
g = \frac{q(1-{\cal P}(0))}{1+p_sq{\cal P}(0)} 
\label{eq8}
\end{equation}
and ${\cal P}(0)$ is given as the root in the interval $(0,1)$ of the cubic 
equation 
\begin{eqnarray}
cp_s^2q^2{\cal P}(0)^3 &+& q[qp_s^2(1-2c) + p_s(1+c) +c]
{\cal P}(0)^2\nonumber\\
&+&[qp_s(1-3c)-2qc +1]{\cal P}(0)  - pc \, = \, 0 . 
\label{eq9}
\end{eqnarray}

In the special limit $p_s \rightarrow 0$, 
$$
{\tilde{\cal P}}(1) \rightarrow \frac{qy}{c}\left(
1-\frac{y}{c}\right),
$$ 
$$
g \rightarrow \frac{qy}{c},
$$
and 
$$
z \rightarrow \frac{(py/c)}{1-(qy/c)} = 1 - \frac{y}{1-c} 
$$
where 
\begin{equation}
y = \frac{1}{2q}\left[1 - \sqrt{1 - 4qc(1-c)}\right]  
\label{eq10}
\end{equation}
is the solution of the quadratic equation 
\begin{equation}
qy^2 - y +c(1-c) = 0.
\label{eq11}
\end{equation}
Moreover, in the special case $p_s$ = 0 , the cubic equation (\ref{eq9}) 
for ${\cal P}_{dh}(0)$ reduces to the quadratic equation 
$$
qc{\cal P}_{dh}(0)^2+(1-2qc){\cal P}_{dh}(0) - pc = 0,
$$
whose physically relevant solution 
$$
{\cal P}_{dh}(0) = \frac{(2qc-1)+ \sqrt{(2qc-1)^2+4pqc^2} }{2qc} 
= 1-\frac{y}{c}$$ 
is identical with the corresponding exact result
\cite{Schadschneider97a,Chowdhury97} for the NaSch model.  
Thus, in the limit $p_s \rightarrow 0$, 
$$
{\cal P}_{dh}(1) = {\cal P}(1) + {\tilde{\cal P}}(1) \rightarrow 
\frac{y^2}{c(1-c)}
$$ 
and
$$
{\cal P}_{dh}(k) \rightarrow \frac{y^2}{c(1-c)}\cdot \biggl[
1 - \frac{y}{1-c}\biggr]^{k-1} \quad {\rm for\ } k \geq 1, 
$$ 
which are the corresponding known results for the NaSch model
\cite{Schadschneider97a,Chowdhury97}.  

\begin{figure}[!h]
 \centerline{\psfig{figure=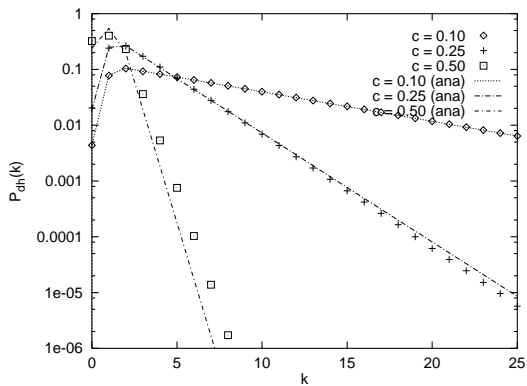,bbllx=50pt,bblly=50pt,bburx=550pt,bbury=400pt,height=5cm}}
\caption{\protect{The DH distributions in the BJH model with 
$V_{max} = 1$ are shown on a semi-log plot for three different 
values of the density of vehicles $c$. The lines are obtained from 
the approximate analytical expressions for ${\cal P}_{dh}(k)$ while 
the discrete data points have been obtained from computer simulation. 
The common parameters are $p = 0.05, p_s = 0.50$.}}
\label{fig1a}
\end{figure}

For $V_{max} = 1$, and for given $c$, we have solved the cubic equation 
(\ref{eq9}) numerically for ${\cal P}(0)$ and, hence, computed 
$g, z, {\tilde{\cal P}}(1), {\cal P}(1), {\cal N}$, ${\cal P}(k)$ for all 
$k \geq 2$. Then, we calculated the distance-headway ${\cal P}_{dh}(k)$
for all $k$ in the BJH model  
corresponding to the density $c$. A comparison of these results
with those obtained from computer simulation for the same set of 
parameters (fig.~\ref{fig1a}) establishes that for low densities,
e.g. $c=0.10$, and $c=0.25$, the approximate analytical expressions 
for ${\cal P}_{dh}(k)$ derived in the COMF approximation, are in good
agreement with the simulation data. Larger deviations of the
analytical results are observable for $c=0.5$. 
%
\begin{figure}[!h]
 \centerline{\psfig{figure=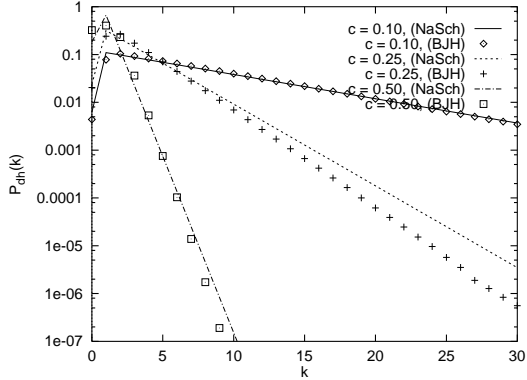,bbllx=50pt,bblly=50pt,bburx=550pt,bbury=400pt,height=5cm}}
\caption{\protect{The DH distributions in the BJH model and 
the NaSch model, both with $V_{max} = 1$ and $p = 0.05$, are shown 
on a semi-log plot for three different values of the density of 
vehicles $c$. The lines are obtained from the exact analytical 
expressions for ${\cal P}_{dh}(k)$ in the NaSch model while the 
discrete data points have been obtained from computer simulation 
of the BJH model with $p_s = 0.50$.}}
\label{fig1b}
\end{figure}

We have made a comparison of the DH distributions in the NaSch model 
and the BJH model, both with $V_{max} = 1$ and with $p = 0.05$, in 
fig.~\ref{fig1b}. At the low density of $c = 0.1$ there is practically no 
difference in the DH distributions in these two models when 
$p = 0.05$. However, at the moderately high density of $c = 0.25$, 
large DH are slightly less probable in the BJH than in the NaSch model. 
Surprisingly, the difference in the DH distributions in these two 
models is less significant at $c = 0.5$ than at $c = 0.25$. This, 
we believe, is a consequence of the fact that at a high density of 
$c = 0.5$ the DH distribution is decided mainly by the blockage 
of one vehicle by another, rather than the slow-to-start rule. 
\begin{figure}[!h]
 \centerline{\psfig{figure=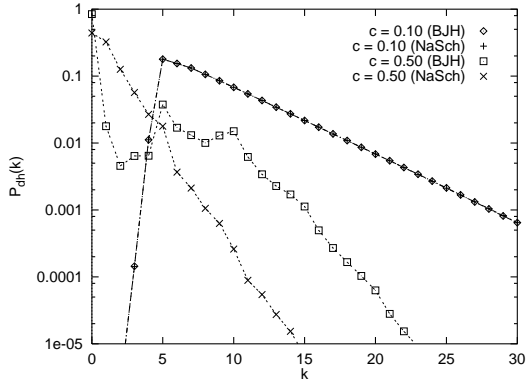,bbllx=50pt,bblly=50pt,bburx=550pt,bbury=400pt,height=5cm}}
\caption{\protect{Same as fig.\ref{fig1a}, except that $V_{max} = 5$ 
and the lines are merely guides to the eye.}}
\label{fig2}
\end{figure}
Some typical DH distributions in the BJH model with $V_{max} = 5$ 
are shown in fig.~\ref{fig2}. The common feature of the DH distributions in 
figs.~\ref{fig1a},\ref{fig1b}, and \ref{fig2} is that for $c = 0.1$, which
corresponds to free-flowing  
traffic, DH smaller than $V_{max}$ are strongly suppressed, while 
beyond $V_{max}$ the DH distribution decays exponentially. The density 
$c = 0.5$ is sufficiently high so that there is a very high probability 
for a vehicle to be stuck in a jam; this is consistent with the maximum 
of $P_{dh}(k)$ located at $k=0$. Similar peaks at $k = 0$ have also 
been observed earlier \cite{Chowdhury97} in the DH distribution in the NaSch model. 
What is more interesting is that the DH distribution in the BJH model 
with $V_{max} = 5$ exhibits peaks also at $k = 5$ and $k = 10$, i.e., 
at $V_{max}$, and $2V_{max}$ (albeit of gradually smaller heights). 
This phenomenon can be understood by considering $p=0$ and recalling 
that, for the values of the parameters used in our simulation, phase 
separation of traffic into a "free-flowing" region and a region of 
jammed vehicles occurs at $c = 0.5$. If $p = 0$, then, the distances 
between the vehicles in the free flowing region of traffic are 
determined by the waiting times of the leading vehicle in a jam. If 
the leading vehicles leaves the jam in the very first opportunity 
then its DH is likely to be $V_{max}$ whereas its likely DH is 
$2 V_{max}$ if it starts at the next time step because of the 
"slow-to-start" rule. This perfect arrangement of the vehicles would 
correspond to delta function-like peaks at $V_{max}$ and $2V_{max}$, 
etc. but gets smeared out to sharp peaks, due to the fluctuations, if $p$ 
is nonzero but not too large. 

\section{\bf The distribution of jam-sizes}

Within the framework of the COMF theory, the distribution of the sizes 
of the jams in the BJH model with $V_{max} = 1$ can be written as
\cite{Schadschneider98a} 
$$
{\cal P}_{js}(k) = {\cal N}_{js} [1-{\cal P}(0)][{\cal P}(0)]^{k-1}
[1-{\cal P}(0)]
$$
where ${\cal N}_{js}$ is a normalization coefficient. Since, the condition 
$\sum_k {\cal P}_{js}(k) = 1$ of normalization leads to 
${\cal N}_{js} = [1-{\cal P}(0)]^{-1}$, the normalized distribution of 
the jam sizes is given by 
\begin{eqnarray}
{\cal P}_{js}(k) = [1-{\cal P}(0)][{\cal P}(0)]^{k-1}
\label{eq12} 
\end{eqnarray} 
In the limit $p_s \rightarrow 0$, 
$$
{\cal P}_{js}(k) \rightarrow \frac{y}{c}\left(1-\frac{y}{c}\right)^{k-1}
$$
which is identical to the jam-size distribution in the NaSch model with 
$V_{max} = 1$ where $y$ is given by the equation (\ref{eq10}). 

\begin{figure}[!h]
 \centerline{\psfig{figure=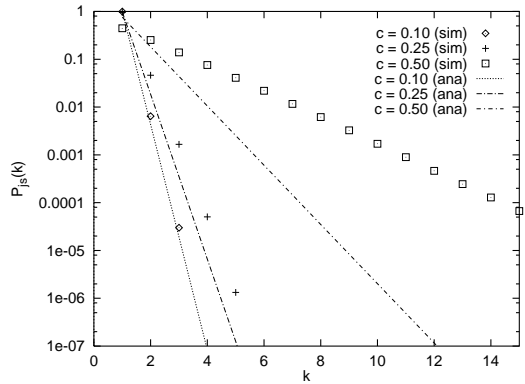,bbllx=50pt,bblly=50pt,bburx=550pt,bbury=400pt,height=5cm}}
\caption{\protect{The distribution of jam-sizes in the BJH 
model with $V_{max} = 1$ are shown on a semi-log plot for the same 
values of $c$, $p$ and $p_s$ as in fig.~\ref{fig1a},\ref{fig1b}. The lines
are obtained  
from the approximate analytical expression (\ref{eq12}) 
for ${\cal P}_{js}(k)$ while the discrete data points have been obtained 
from computer simulation.}}
\label{fig3a}
\end{figure}
The jam-size distribution for the BJH model is well approximated by the 
expression (\ref{eq12}) at low densities of the vehicles as is evident 
from the fig.~\ref{fig3a} where comparison has been made with the
corresponding  
numerical data obtained from computer simulation of the BJH model with 
the same set of parameters. However, the approximate expression 
(\ref{eq12}) deviates more and more from the computer simulation data 
with increasing density for the densities we took into account. 
\begin{figure}[!h]
 \centerline{\psfig{figure=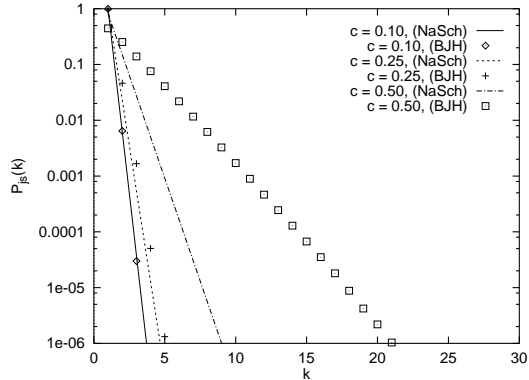,bbllx=50pt,bblly=50pt,bburx=550pt,bbury=400pt,height=5cm}}
\caption{\protect{The distributions of jam-sizes in the BJH 
model and the NaSch model, both with $V_{max} = 1$ and $p = 0.05$, 
are shown on a semi-log plot for three different values of the 
density of vehicles $c$. The lines are obtained from the exact 
analytical expressions for ${\cal P}_{js}(k)$ in the NaSch model 
while the discrete data points have been obtained from computer 
simulation of the BJH model with $p_s = 0.50$.}}
\label{fig3b}
\end{figure}
A comparison between the jam-size distributions in the NaSch and BJH 
models, both for $V_{max} = 1$ and $p = 0.05$, is made in fig.~\ref{fig3b}.
This comparison establishes that the difference in the jam-size 
distributions in these two models becomes wider with the increase 
of vehicle density $c$. At a high density of $c = 0.5$, for example, 
long jams are much more probable in the BJH model than in the NaSch 
model.
\begin{figure}[!h]
 \centerline{\psfig{figure=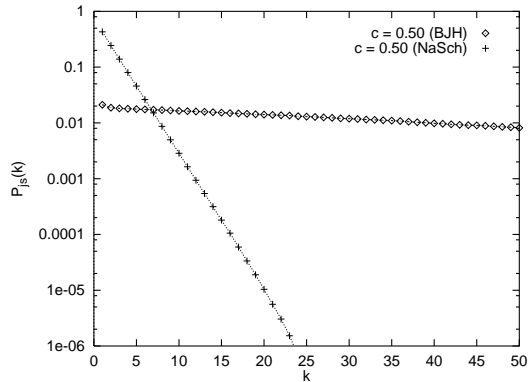,bbllx=50pt,bblly=50pt,bburx=550pt,bbury=400pt,height=5cm}}
\caption{\protect{Same as fig.~\ref{fig3a}, except that $V_{max} = 5$
and the lines are merely guides to the eye.}}
\label{fig4}
\end{figure}
A few typical distributions ${\cal P}_{js}(k)$ for $V_{max} = 5$ are 
shown in fig.~\ref{fig4}. From figs.~\ref{fig3a}, \ref{fig3b} and \ref{fig4}
we conclude that, for both $V_{max} =1$ 
and $V_{max} = 5$, the jam size distributions ${\cal P}_{js}(k)$ decay 
exponentially with the jam-size $k$.

\section{The distribution of gaps between successive jams}

\subsection{A hybrid approach} 

Let us use a {\it three-state} variable $s$ to describe the state 
of a site; $s = -1$ if the site is empty, $s = 0$ if the site is 
occupied by a vehicle with instantaneous speed $V = 0$ and 
$s = 1$ if the site is occupied by a vehicle with instantaneous 
speed $V = 1$ \cite{Chowdhury97}. Note that this variable $s$ does not
distinguish  
between the slow-to-start vehicles and other vehicles which have 
instantaneous speed $V = 0$. We denote the probability of a 
configuration immediately after the slow-to-start stage by the 
symbol $P(a,b \cdots )$ while that immediately after the randomization 
stage by $p(c,d \cdots )$ and use the sequence $5-1-2-3-4$ of the steps 
of the update rules, instead of $1-2-3-4-5$. 

In terms of the state variable $s$, a $2-cluster-like$ approximate 
decomposition of the probability distribution of the gaps between 
successive jams leads to the expression 
\begin{equation}
{\cal P}_{jg}(k) = {\cal N}_{jg} \biggl[\sum_{\{s_i=\pm 1\}} 
p_2(\underline{0}\mid s_1)
  p_2(\underline{s_1}\mid s_2)\cdots p_2(\underline{s_{k-1}}\mid s_k)
 p_2(\underline{s_k}\mid 0)\biggr] \quad {\rm for\ } k \geq 1 
\label{eq13} 
\end{equation} 
where ${\cal N}_{jg}$ is a normalization coefficient and the conditional 
probabilities $p_2(\underline{a}|b)$ are defined through the equation  
\begin{equation}
p_2(\underline{a}|b) = \frac{p_2(a,b)}{p_2(a,0)+p_2(a,-1)+p_2(a,1)}. 
\label{eq14} 
\end{equation}

Let us now define the $2 \times 2$ transfer matrix $T$ with the 
following elements:\\ 
\begin{equation}\label{eq15}
\begin{aligned}
T[1,1]&=p_2(\underline{1}\mid 1),\qquad\\ 
T[2,1]&= p_2(\underline{-1}\mid 1),\qquad
\end{aligned}
\begin{aligned}
T[1,2]&= p_2(\underline{1}\mid -1),\\
T[2,2]&= p_2(\underline{-1}\mid -1). 
\end{aligned}
\end{equation}
In terms of this transfer matrix, the expression (\ref{eq13}) for 
$P_{jg}(k)$ can be recast as  
\begin{eqnarray}
{\cal P}_{jg}(k) &=& {\cal N}_{jg} \bigl[p_2(\underline{0}\mid 1)
T^{k-1}[1,1]p_2(\underline{1}\mid 0)  
+ p_2(\underline{0}\mid -1)T^{k-1}[2,1]p_2(\underline{1}\mid 0) \nonumber \\ 
&+& p_2(\underline{0}\mid 1)T^{k-1}[1,2]p_2(\underline{-1}\mid 0) +
 p_2(\underline{0}\mid -1)T^{k-1}[2,2]p_2(\underline{-1}\mid 0)\bigr] 
\label{eq16} 
\end{eqnarray}
where $T^{k-1}[i,j]$ refers to the [i,j] element of the matrix $T^{k-1}$.  
It is straightforward to show that  
\begin{eqnarray}
T^{k-1}[1,1] &=& (\lambda_1 \lambda_2^{k-1}-\lambda_2 \lambda_1^{k-1})/
 (\lambda_1 - \lambda_2), 
\label{eq17}\\
T^{k-1}[1,2] &=& (\lambda_1^{k-1}- \lambda_2^{k-1})/(\lambda_1 - \lambda_2), 
\label{eq18}\\
T^{k-1}[2,1] &=& (\lambda_1 \lambda_2^{k}-\lambda_2 \lambda_1^{k})/(\lambda_1 
- \lambda_2), \label{eq19}\\
T^{k-1}[2,2] &=& (\lambda_1^{k}- \lambda_2^{k})/(\lambda_1 - \lambda_2), 
\label{eq20}
\end{eqnarray} 
where $\lambda_1$ and $\lambda_2$ are the two eigenvalues of the transfer 
matrix $T$.

As shown in the appendix A, the 2-cluster-like conditional probabilities 
$p_2(\underline{a}|b)$ can be related to the quantities like $g, z, 
{\cal P}(0), {\cal P}(1), {\tilde{\cal P}}(1)$, etc. which occur in 
the COMF theory. Using these relations we get  
\begin{eqnarray}
T[1,1] &=& p_2(\underline{1}\mid 1) =0,
\label{eq21}\\
T[1,2] &=& p_2(\underline{1}\mid -1) = 1,
\label{eq22}
\end{eqnarray}
\begin{eqnarray}
T[2,1] &=& p_2(\underline{-1}\mid 1) = \frac{cg(1-{\cal P}(0))}{c[1-
{\cal P}(0)]+z(1-c)},
\label{eq23}\\
T[2,2] &=& p_2(\underline{-1}\mid -1) = \frac{z(1-c)}{c[1-{\cal P}(0)]
+z(1-c)},\label{eq24} 
\end{eqnarray}
and, hence \cite{Sinha98}, with $D = c[p + q{\cal P}(0) 
+ q p_s{\tilde{\cal P}}(1)]$, 
\begin{eqnarray}
{\cal P}_{jg}(k)&=& {\cal N}_{jg}\cdot \frac{c{\cal P}(0)g}{D}\cdot
\frac{ \lambda_1^{k-1}- \lambda_2^{k-1}}{\lambda_1 - \lambda_2}\cdot
\frac{c(1-g)[1-{\cal P}(0)]}{c[1-{\cal P}(0)]+z(1-c)} \nonumber \\
&+& \left(1-\frac{c{\cal P}(0)}{D}\right)\cdot\frac{ \lambda_1^{k}- 
\lambda_2^{k}}{\lambda_1 - \lambda_2}\cdot \frac{c(1-g)
[1-{\cal P}(0)]}{c[1-{\cal P}(0)]+z(1-c)}
\label{eq25} 
\end{eqnarray}
where 
\begin{eqnarray}
\lambda_{1,2} & = & \frac{1}{2}\Biggl[\frac{z(1-c)}{z(1-c)+c[1-{\cal P}(0)]}
 \nonumber \\
& \pm & \sqrt{\biggl(\frac{z(1-c)}{z(1-c)+c[1-{\cal P}(0)]}\biggr)^2+ 
\frac{4cg[1-{\cal P}(0)]}{z(1-c)+c[1-{\cal P}(0)]}}\Biggr] 
\label{eq26} 
\end{eqnarray}
are the eigenvalues of the $T$ matrix in this approximation. 
From equations (\ref{eq25}) and (\ref{eq26}) we find that, 
in the limit $p_s \rightarrow 0$, 
$$\lambda_{1,2} \rightarrow \frac{1}{2}\left[\left(1 - \frac{y}{1-c}
\right)
\pm \sqrt{\left(1 - \frac{y}{1-c}\right)^2 + 4 \left(\frac{y}{c(1-c)}-1
\right)}\right], $$
and 
$$
{\cal P}_{jg}(k) \rightarrow \frac{py^2c(\lambda_1^k-\lambda_2^k)
+qy^2(c-y)(\lambda_1^{k-1}-\lambda_2^{k-1})}{c^2(1-c)(\lambda_1-\lambda_2)} 
$$
which is the jam-gap distribution in the NaSch model for $V_{max} = 1$ 
\cite{Chowdhury97}. 

The probabilities like $p_2(s_{k-1},s_k)$ which appeared in the 
intermediate steps of the calculations in this subsection are not 
true 2-cluster probabilities in the usual sense of the term
\cite{Schreckenberg95}.   
The final expression for the distribution $P_{jg}(k)$ involves only 
quantities which we compute entirely within the framework of the 
COMF theory. Therefore, this whole approach may be called a {\it 
hybrid} approach. We shall compare the expression (\ref{eq25}) for 
${\cal P}_{jg}(k)$ as well as the corresponding expression derived 
in the next subsection following the true 2-cluster approach with 
the data obtained from computer simulation at the end of this section.

\subsection{The 2-cluster approach}

We now derive the distribution $P_{jg}(k)$ using the true 2-cluster 
approximation. Let us now allow the state variable $s$ to take one 
of the four possible values, namely, $s = -1$ corresponding to an 
empty site, $s = 1$ corresponding to a site occupied by a vehicle 
with instantaneous speed $V = 1$ while $s = 0$ and $2$ correspond 
to a site occupied by, respectively, a non-slow-to-start vehicle 
with $V = 0$ and a slow-to-start vehicle (also with $V = 0$). 
In terms of this new definition of the variable $s$, we can write 
the distribution of the gaps between the jams in the BJH model with 
$V_{max} = 1$ as 
\begin{equation}
{\cal P}_{jg}(k) = {\cal N}'_{jg} \sum_{\sigma,\sigma'=0,2} 
\sum_{\{s_i=\pm 1\}} 
C_2(\underline{\sigma}\mid s_1)
  C_2(\underline{s_1}\mid s_2)\cdots C_2(\underline{s_{k-1}}\mid s_k)
 C_2(\underline{s_k}\mid \sigma') 
\label{eq27} 
\end{equation} 
where ${\cal N}'_{jg}$ is a normalization coefficient and 
$C_2(\underbar{a}|b)$ are the true conditional 2-cluster 
probabilities. In terms of the transfer matrix $T$, whose elements 
are given by the equation 
\begin{equation}\label{eq28}
\begin{aligned}
T[1,1]&=C_2(\underline{1}\mid 1),\qquad\\
T[-1,1]&=C_2(\underline{-1}\mid 1),\qquad 
\end{aligned} 
\begin{aligned}
T[1,-1]&= C_2(\underline{1}\mid -1),\\
T[-1,-1]&= C_2(\underline{-1}\mid -1),
\end{aligned} 
\end{equation}
the equation (\ref{eq27}) can be written as 
\begin{equation}
{\cal P}_{jg}(k) =  {\cal N}'_{jg} \sum_{\{\sigma,\sigma'=0,2\}} 
\sum_{s,s'=\pm 1} C_2(\underline{\sigma}\mid s)T^{k-1}[s,s']
C_2(\underline{s'}|\sigma').
\label{eq29} 
\end{equation}
For convenience we have labeled the elements of $T$ by (1,1), (1,-1), (-1,1),
(-1,-1) instead of (1,1), (1,2), (2,1), (2,2).
As before, $T^{k-1}[i,j]$ refers to the $[i,j]$ element of the 
matrix $T^{k-1}$ and are given by the equations (\ref{eq17}-\ref{eq20}).
We compute the 2-cluster probabilities as outlined in appendix B 
and substituting in equation (\ref{eq29}) get the distribution of 
gaps between successive jams in the BJH model with $V_{max} = 1$.
\begin{figure}[!h]
 \centerline{\psfig{figure=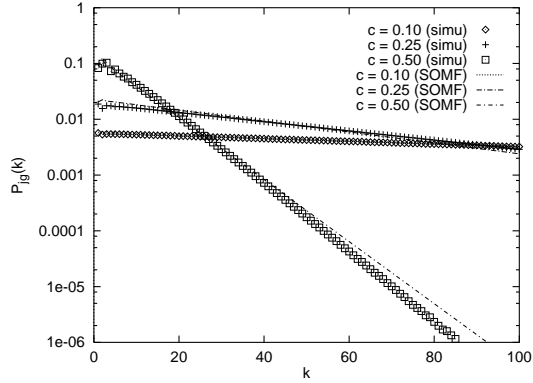,bbllx=50pt,bblly=50pt,bburx=550pt,bbury=400pt,height=5cm}}
\caption{\protect{The distribution of sizes of gaps between 
successive jams in the BJH model with $V_{max} = 1$ are shown on 
a semi-log plot for the same values of $c$, $p$ and $p_s$ as in
fig. \ref{fig1a},\ref{fig1b}. 
The lines are obtained from the approximate analytical expression 
(\ref{eq29}) for ${\cal P}_{jg}(k)$ while the discrete data points 
have been obtained from computer simulation.}}
\label{fig5a}
\end{figure}
\begin{figure}[!h]
 \centerline{\psfig{figure=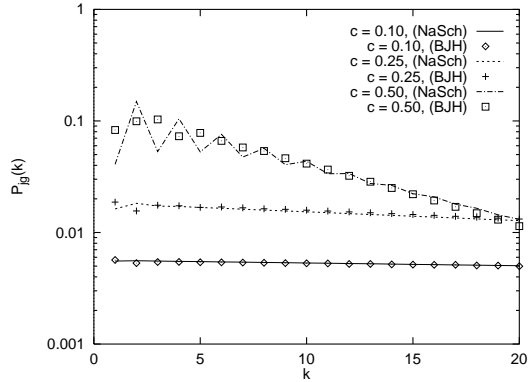,bbllx=50pt,bblly=50pt,bburx=550pt,bbury=400pt,height=5cm}}
\caption{\protect{The distributions of the gaps between 
successive jams in the BJH model and the NaSch model, both with 
$V_{max} = 1$ and $p = 0.05$, are shown on a semi-log plot for 
three different values of the density of vehicles $c$. The lines 
are obtained from the exact analytical expressions for 
${\cal P}_{jg}(k)$ in the NaSch model while the discrete data points 
have been obtained from computer simulation of the BJH model with 
$p_s = 0.50$.}}
\label{fig5b}
\end{figure}
Just as in the case of the jam-size distribution, the agreement of 
expressions (\ref{eq25}) and (\ref{eq29}) with the corresponding 
computer simulation data is better when the vehicle density is lower 
(fig.~\ref{fig5a}). We have found that, at lower densities, the prediction of 
the hybrid approach (i.e., the equation (\ref{eq25})), rather than 
that of the true 2-cluster approximation (i.e., the equation 
(\ref{eq29}), is in better agreement with the corresponding 
computer simulation data. However, beyond some intermediate density, 
the results of the true 2-cluster approximation are found to be  
closer to the computer simulation data than the corresponding 
results obtained in the hybrid approach. It is interesting to note 
that, for the values of the set of parameters chosen for fig.~\ref{fig5a}, 
there is very little difference between the jam-gap distributions in 
the NaSch and BJH models (see fig.~\ref{fig5b}). 

\begin{figure}[!h]
 \centerline{\psfig{figure=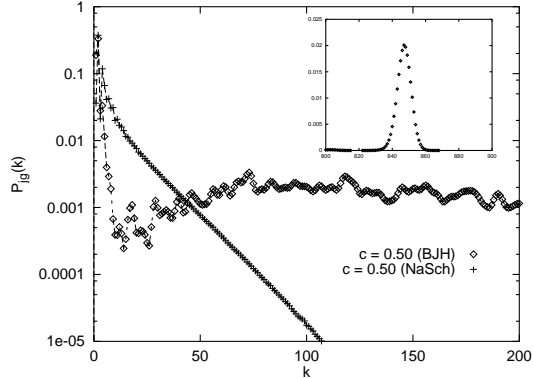,bbllx=50pt,bblly=50pt,bburx=550pt,bbury=400pt,height=5cm}}
\caption{\protect{Same as fig.~\ref{fig5a}, except that $V_{max} = 5$
and the lines are merely guides to the eye. The inset shows the local
maximum at $\bar{L}_f(c)$.}}
\label{fig6}
\end{figure}

Comparing the curves in 
fig.~\ref{fig5a},\ref{fig5b} with the corresponding distributions for
$V_{max} = 5$, shown  
in fig.~\ref{fig6}, we conclude that the distribution  ${\cal P}_{jg}(k)$ 
decays exponentially with the gap $k$ between the successive jams 
for all finite $V_{max}$, as far as small distances are
considered. Nevertheless ${\cal P}_{jg}(k)$ allows to identify
macroscopic free flow regimes. In particular, we found for $v_{max}
=5$, $p_s =0.5$, and $p=0.05$ a local maximum of the distance distribution at
$\bar{L}_f(c)=L-\bar{L}_j(c)$, where $\bar{L}_f(c)$ denotes the
average size of the free flow regime at a given density, and
$\bar{L}_f(c)$ the
typical size of the macroscopic jam. Therefore the jam-gap
distribution is well suited to identify at least the macroscopic free flow
regime of a phase separated high density state.

\section{The TH distribution} 

We label the position of the detector by $j = 0$, the site immediately 
in front of it by $j = 1$, and so on. For the convenience of analytical 
calculations of the TH distribution in the steady state of the BJH  
model with $V_{max} = 1$, we assume the sequence of steps $2-3-4-5-1$, 
just as in the corresponding calculation for the NaSch model 
\cite{Ghosh98,Chowdhury98} the sequence of steps was assumed to be
$2-3-4-1$.   

Suppose, ${\cal P}_m(t_1)$ is the probability that the FV takes time 
$t_1$ to reach the detector moving from its initial position, where it 
was located when the clock at the detector site was set to $t = 0$, 
after the LV just left the detector site. Suppose, after reaching the 
detector site, the FV waits there for $\tau - t_1$ time steps, either 
because of the presence of another vehicle in front of it or because 
of its own random braking; the probability for this event is denoted by 
$Q(\tau-t_1|t_1)$. Then, the distribution ${\cal P}_{th}(\tau)$, of the 
TH $\tau$, is given by \cite{Ghosh98} 
$$
{\cal P}_{th}(\tau) = \sum_{t_1=1}^{\tau-1} {\cal P}_m(t_1) 
Q(\tau-t_1|t_1)
$$ 
or, equivalently, by 
\begin{equation} 
{\cal P}_{th}(\tau) = \sum_{t_w=1}^{\tau-1} {\cal P}_m(\tau-t_w) 
Q(t_w|\tau-t_w) 
\label{eq30}
\end{equation} 
where $t_w$ is the waiting time at the detector site.

We calculate ${\cal P}_{th}(\tau)$ for $V_{max} = 1$ analytically using 
the equation (\ref{eq30}). In order to calculate ${\cal P}_m(t_1)$ we 
consider those spatial configurations at $t = 0$ from which the FV can 
reach the detector site within $t_1$ time steps. This implies that one 
needs to consider configurations up to a maximum separation of $t_1$ 
sites from the detector site so that the FV vehicle can reach the 
detector site even from the farthest point with $t_1$ hops. Thus, the 
configurations of interest are of the form 
$$(\underbrace{1,-1,....,-1}_{n~ times}|\underbar{-1})$$ 
where $n = 1,2,...,t_1$. The underlined $-1$ implies that we have 
to find the conditional steady-state probability for the given 
configuration subject to the condition that the underlined site 
(detector site) has just become empty. In the 2-cluster approximation, 
this probability, $\Pi(n)$, is given by 
\begin{equation}
\Pi(n) = P_2(1|\underbar{-1})\{P_2(-1|\underbar{-1})\}^{n-1}. 
\label{eq31}
\end{equation}
where, 
the 2-cluster steady-state probabilities are given by 
\begin{equation}
P_2(1|\underbar{-1}) = \frac{c}{d}[1-{\cal P}_0-p_s{\tilde{\cal P}}(1)]
\label{eq32}
\end{equation}
and  
\begin{equation}
P_2(-1|\underbar{-1}) = z = \frac{pg}{q{\bar{g}}} 
\label{eq33}
\end{equation}

For all configurations with $t_1 > n$, the vehicle has to stop 
$(t_1-n)$ times in order to reach the detector site after exactly 
$t_1$ time steps. Equivalently, $t_1 - 1$ time steps must elapse 
in crossing $n_1$ bonds (as the vehicle must certainly hop across 
the bond connecting the detector site with its preceding site at 
the last time step); the number of ways in which $t_1-1$ time 
steps can be distributed among $n-1$ bonds is ${{t_1-1} \choose {n-1}}$. 
Thus,
\begin{equation}
{\cal P}_m(t_1) = \sum_{n=1}^{t_1} ~{{t_1-1}\choose {n-1}} 
\Pi(n) q^n p^{t_1-n} 
\label{eq34}
\end{equation}
as a factor $q$ for every forward step and a factor of $p$ for 
every halt must be included. Substituting (\ref{eq32}) and (\ref{eq33}) 
into equation (\ref{eq31}) and using the corresponding expression for 
$\Pi(n)$ in (\ref{eq34}) we get 
\begin{equation}
{\cal P}_m(t_1) = {\cal N}_t \left(1 - {\cal P}_0 - p_s{\tilde{\cal P}}(1)
\right)\frac{qc}{d}\left(\frac{p}{\bar{g}}\right)^{t_1-1}
\label{eq35}
\end{equation}
where ${\cal N}_t$ is the normalization factor. The normalization  
condition $ \sum_{t_1=1}^{\infty} {\cal P}_m(t_1) = 1$ leads to 
$${\cal P}_m(t_1) = \frac{\bar{g}-p}{\bar{g}}\left(\frac{p}{\bar{g}}
\right)^{t_1-1}$$
or, equivalently, 
\begin{eqnarray}
{\cal P}_m(\tau-t_w) = \frac{\bar{g}-p}{\bar{g}}
\left[1 - \frac{\bar{g}-p}{\bar{g}} \right]^{\tau-t_w-1}
\label{eq36}
\end{eqnarray}
It is straightforward to verify that in the limit $p_s \rightarrow 0$, 
$(\bar{g}-p)/\bar{g} \rightarrow (qy/d)$ and, hence, from the expression 
(\ref{eq36}) we find that  
$$
{\cal P}_m(t_1) \rightarrow \frac{qy}{d}\left[1 - \frac{qy}{d}
\right]^{t_1-1}
$$
which is the corresponding expression for the NaSch model \cite{Ghosh98}. 

Next, let us derive an analytical expression for $Q(t_w|\tau-t_w)$. 
When the FV arrives at the detector site exactly $\tau-t_w$ time 
steps after the departure of the LV, the LV can be at any of the 
sites labeled by $1, 2, ...\tau-t_w+1$. We consider two situations 
separately, namely, (i) the LV is not at site '1' when the FV arrives 
at the detector site, and (ii) the LV is still at site '1' when the 
FV arrives at the detector site. Let us first consider the situation 
(i). The probability that the LV stays at the site '1', i.e., it does 
not move in any of the $\tau - t_w$ time steps = $(\bar{g})^{\tau-t_w}$. 
Therefore, the probability that the LV is not at site '1' 
$= 1 - (\bar{g})^{\tau-t_w}$. If the LV is not at site '1' then the 
probability that the FV halts at the detector site for exactly 
$t_w$ time steps is $p^{t_w-1}q$ because it should halt due to 
random braking for exactly $t_w-1$ time steps and move at the last 
time step. Thus, the contribution to $Q(t_w|\tau-t_w)$ coming from 
all those situations where the LV is not at site '1' when the FV 
reaches the detector site is 
$$ 
\left[1 - (\bar{g})^{\tau-t_w}\right] p^{t_w-1}q  \eqno (I)
$$ 

Let us now calculate the contribution to $Q(t_w|\tau-t_w)$ from 
those situations which are of type (ii), i.e., where the LV is at 
site '1' when the FV reaches the detector site. So, the LV will 
have to move so that the FV is able to leave the detector site 
(and move forward) after $t_w$ time steps. Suppose, the LV moves 
from '1' after $k$ time steps ($1 \leq k \leq t_w-1$), then the 
FV will have to stay at the detector site for the next $t_w-1-k$ 
time steps either because of random braking or because of the 
slow-to-start rule and, then, move forward at the last time step. 
Note that if $t_w = 1$, i.e., if the FV stays at the detector site 
only for one time step, then the site '1' must be empty when the 
FV reaches the detector site; the corresponding contribution to 
$Q(t_w|\tau-t_w)$ has already been included in (I). The calculation 
of $Q(t_w|\tau-t_w)$ for all $\tau \geq 2$ is quite non-trivial. 
Suppose, the LV moves from the site '1' after $k$ time steps and 
$k < t_w-1$. Then the FV can stay at the detector site during the 
immediately next time step with probability $p_s + q_sp$ and for 
the remaining $t_w - 2 - k$ time steps with probability $p$ each; 
in the last step it must move forward with probability $q$. 
There is, however, yet another possibility. When the LV moves from 
'1' after $t_w-1$ time steps, then the FV has to move forward in 
the last time step; the corresponding probability being $q q_s$. 
Thus, the contribution to all these situations of type (ii) is 
$$
(\bar{g})^{\tau-t_w} q g \left[ \sum_{k=1}^{t_w-2} (\bar{g})^{k-1} 
p^{t_w-2-k} (p_s+q_sp) + (\bar{g})^{t_w-2} q_s \right]
$$ 
$$ 
= (\bar{g})^{\tau-t_w} q g \left[ \frac{\alpha(\bar{g})^{t_w-2} - 
\beta p^{t_w-2}}{\bar{g}-p} \right]  \eqno (II) 
$$ 
where $\alpha = p_s + q_s \bar{g}$ and $\beta = p_s + q_s p$. Thus, 
combining (I) and (II) we find 
\begin{equation}
Q(t_w|\tau-t_w) = [1 - (\bar{g})^{\tau-1}] q \quad for \quad t_w = 1 
\label{eq37}
\end{equation}
and 
\begin{eqnarray}
Q(t_w|\tau-t_w) & = & [1 - (\bar{g})^{\tau-t_w}] p^{t_w-1}q \nonumber \\
& + & (\bar{g})^{\tau-t_w} q g \biggl[ \frac{\alpha(\bar{g})^{t_w-2} - 
\beta p^{t_w-2}}{\bar{g}-p} \biggr] \quad {\rm for\ } t_w \geq 2  
\label{eq38}
\end{eqnarray}
In the limit $p_s \rightarrow 0$, 
$$
Q(\tau-t_1|t_1) \rightarrow  [1-(\bar{g})^{t_1}]p^{\tau-t_1-1}q 
\nonumber \\
+ (\bar{g})^{t_1}g q \biggl[\frac{(\bar{g})^{\tau-t_1-1}-(p)^{\tau-t_1-1}
}{\bar{g}-p}\biggr]
$$
which is, indeed, the corresponding expression for $Q(\tau-t_1|t_1)$ 
in the NaSch model \cite{Chowdhury98}.

Finally, using (\ref{eq37}) and (\ref{eq38}) in (\ref{eq30}) and carrying 
out the summation we get 
\begin{eqnarray}
{\cal P}_{th}(\tau) &= &{\cal N}_{th}\biggl[\frac{\bar{g}-p}{\bar{g}}\cdot
\frac{(1/\bar{g})^{\tau-1}-1}{(1/\bar{g})-1}\cdot q p^{\tau-2}  
+ \frac{\alpha qg}{\bar{g}}\cdot \frac{(\bar{g})^{\tau-1}- 
p^{\tau-1}}{\bar{g}-p} \nonumber \\
& - & (\bar{g}-p) q p^{\tau-2}(\tau-1) 
-g q \beta p^{\tau-3} (\tau-1) 
- g q p^{\tau - 2} \left(\frac{\alpha}{\bar{g}} - \frac{\beta}{p}
\right)\biggr] 
\label{eq39}
\end{eqnarray}
where ${\cal N}_{th}$ is the normalization coefficient (which is required 
as the expression (\ref{eq39}) is not exact). From the expression 
(\ref{eq39}) we also find that in the limit $p_s \rightarrow 0$,
$$
{\cal P}_{th}(\tau) = \frac{qy}{c-y}\left[1-\frac{qy}{c}
\right]^{\tau-1} 
+ \frac{qy}{d-y}\left[1-\frac{qy}{d}\right]^{\tau-1}
$$ 
$$
- \left[\frac{qy}{c-y} + \frac{qy}{d-y} \right] p^{\tau-1} 
- q^2 (\tau-1) p^{\tau-2}
$$
which is the known exact expression for ${\cal P}_{th}(\tau)$ in the 
NaSch model \cite{Chowdhury98}. 
\begin{figure}[!h]
 \centerline{\psfig{figure=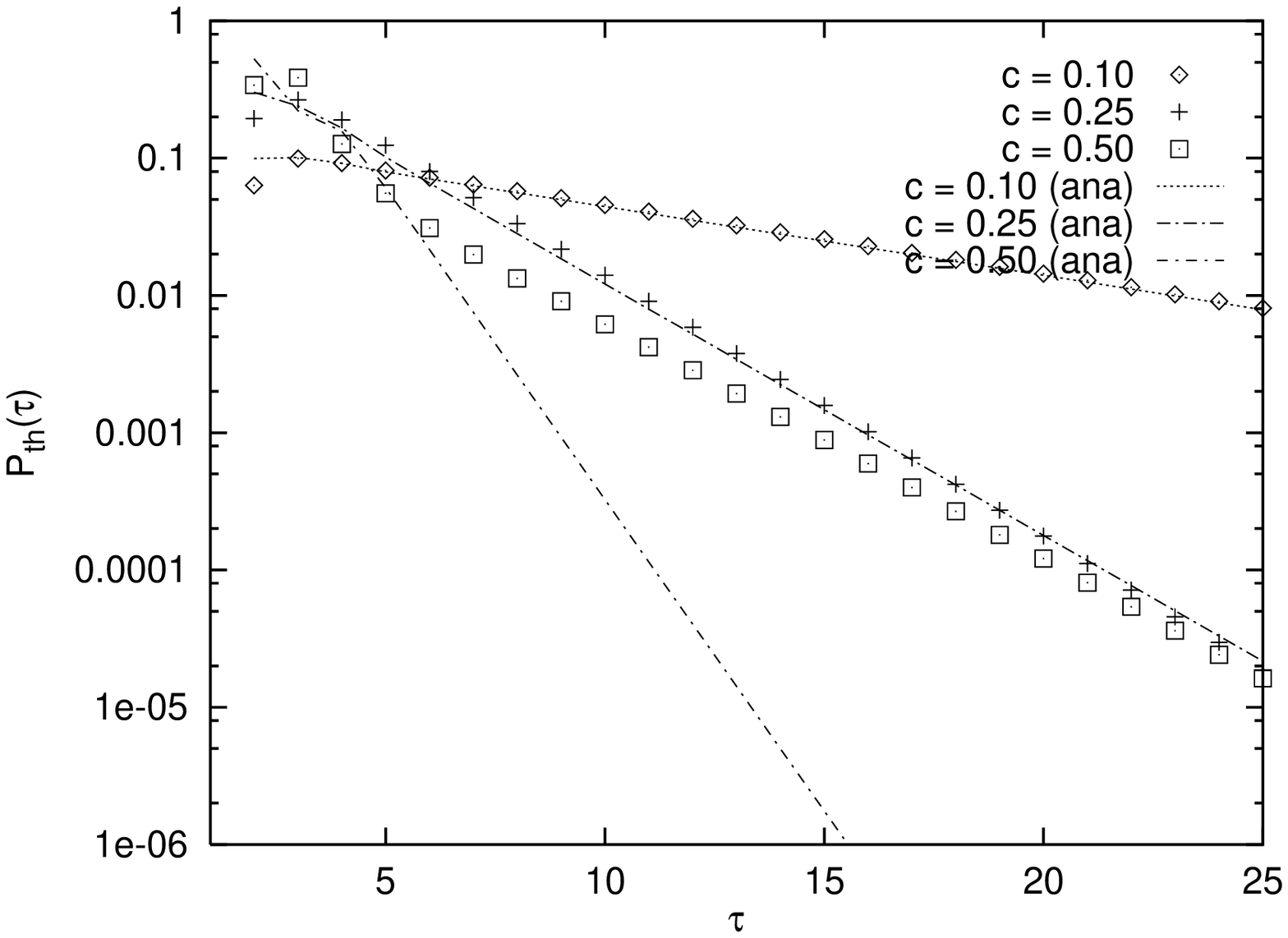,bbllx=50pt,bblly=50pt,bburx=550pt,bbury=400pt,height=5cm}}
\caption{\protect{The TH-distribution in the BJH model with 
$V_{max} = 1$ are shown on a semi-log plot for the same values of 
$c$, $p$ and $p_s$ as in figs.~\ref{fig1a}, \ref{fig1b}. The lines are
obtained from the  
approximate analytical expression (82) for ${\cal P}_{th}(\tau)$ 
while the discrete data points have been obtained from computer 
simulation.}}
\label{fig7a}
\end{figure}
\begin{figure}[!h]
 \centerline{\psfig{figure=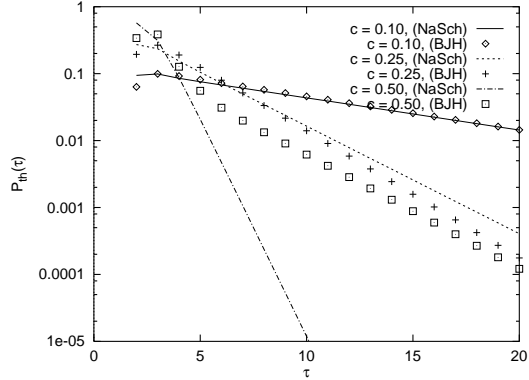,bbllx=50pt,bblly=50pt,bburx=550pt,bbury=400pt,height=5cm}}
\caption{\protect{The TH distributions in the BJH model and 
the NaSch model, both with $V_{max} = 1$ and $p = 0.05$, are shown 
on a semi-log plot for three different values of the density of 
vehicles $c$. The lines are obtained from the exact analytical 
expressions for ${\cal P}_{th}(k)$ in the NaSch model while the 
discrete data points have been obtained from computer simulation 
of the BJH model with $p_s = 0.50$. }}
\label{fig7b}
\end{figure}

The TH-distribution for the BJH model with $V_{max} = 1$ is well 
approximated by the expression (\ref{eq39}) at low densities of the 
vehicles but the approximate expression (82) deviates more and more 
from the computer simulation data with increasing density
(fig.~\ref{fig7a}). A  
comparison of the TH distributions in the NaSch and BJH models, both 
for $V_{max} = 1$ and $p = 0.05$, in fig.~\ref{fig7b}, clearly demonstrates 
that the larger is the density $c$ the wider is the difference 
between the TH distributions in these two models. In fact, at 
a high density, e.g., $c = 0.5$, the larger TH are much more 
probable in the BJH model than in the NaSch model. This is basically
for two reasons. Firstly the slow-to-start rule leads to larger jams
compared to the NaSch model at the same density (see fig.~\ref{fig3b}) and
secondly the downstream velocity of the jams is reduced, which also
leads to an increase of the typical waiting time for a car in a jam. 
\begin{figure}[!h]
 \centerline{\psfig{figure=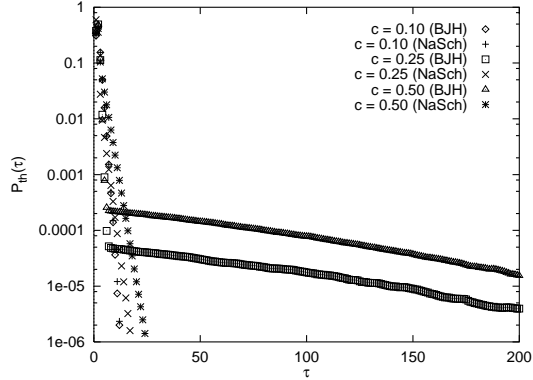,bbllx=50pt,bblly=50pt,bburx=550pt,bbury=400pt,height=5cm}}
\caption{\protect{Same as fig.~\ref{fig7a}, except that $V_{max} = 5$.}}
\label{fig8}
\end{figure}
The computer simulation data for the same densities as in
fig.~\ref{fig7a},\ref{fig7b}, but 
for $V_{max} = 5$ are shown in fig.~\ref{fig8}. From the definition of TH it 
is clear that, for the NaSch model as well as the BJH model, the TH 
distributions must vanish for $\tau < 2$ if $V_{max} = 1$ but need 
not do so if $V_{max} > 1$ \cite{Ghosh98}. Except for this difference, at low 
densities, e.g., $c = 0.1$ the TH distributions ${\cal P}_{th}(\tau)$ 
for $V_{max} = 1$ is qualitatively similar to that for $V_{max} = 5$; 
both exhibit a maximum and fall exponentially with $\tau$ beyond the 
most-probable TH. But, there are some qualitative differences in the 
$\tau$-dependence of ${\cal P}_{th}(\tau)$ for $V_{max} = 1$ and that 
for $V_{max} = 5$ in the BJH model. Beyond the most probable TH, the 
TH distribution for $V_{max} = 1$ can be fitted to a exponentially 
decaying function with a single decay rate. On the other hand, for 
$V_{max} = 5$, there are two regimes of $\tau$: in the small $\tau$ 
regime, which corresponds to the "free-flowing" regions of traffic, 
${\cal P}_{th}(\tau)$ can be fitted to an exponentially decaying 
function (with a fast rate of decay) whereas in the large $\tau$ regime, 
which corresponds to the jammed region of traffic, ${\cal P}_{th}(\tau)$ 
can be fitted also to an exponentially-decaying function (with a very 
slow rate of decay). This phenomenon of two different regimes of $\tau$, 
which arises from the phase separation in the BJH model, is absent in 
the NaSch model \cite{Ghosh98}.   

\section{Regime of validity of the analytical expressions} 

In the preceding section we have observed that the analytical 
expressions derived for the BJH model with $V_{max} = 1$ are is 
good agreement with the numerical data from our computer simulation 
if the density of vehicles is not high. However, the higher is 
the vehicle density the stronger are those correlations which have 
been neglected in our calculations within the COMF approximation 
and the 2-cluster approximation; these correlations lead to increasing 
deviation of the analytical result, from the corresponding simulation 
data, with increasing density of the vehicles. We now make a 
quantitative estimation of the regime of validity of our analytical 
results derived in this paper. 

For the NaSch model, the 2-cluster approximation, which gives exact 
results for $V_{max} = 1$, is known to be equivalent to the 
car-oriented mean-field (COMF) theory, which neglects the correlations 
of the type  
\begin{equation}
G_{n,m} = {\cal P}(n,m) - {\cal P}(n) {\cal P}(m) 
\label{eq40}
\end{equation} 
where ${\cal P}(n,m)$ is the joint probability of finding a DH $m$ 
followed immediately by a DH $n$. On the other hand, the COMF does 
not give exact results for the BJH model \cite{Schadschneider97}
because of non-vanishing  
correlations of this type. 
\begin{figure}[!h]
 \centerline{\psfig{figure=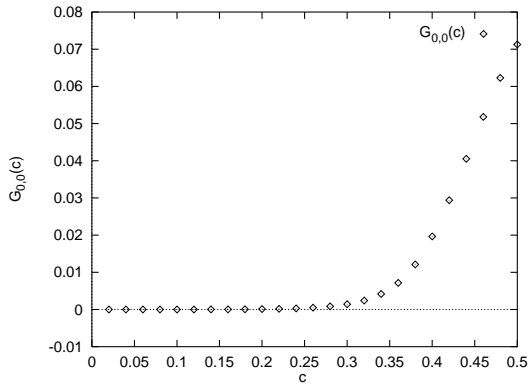,bbllx=50pt,bblly=50pt,bburx=550pt,bbury=400pt,height=5cm}}
\caption{\protect{The correlation function $G_{0,0}(c)$ 
in the BJH model with $V_{max} = 1$ is plotted against $c$ for the 
same values of the parameters $p$ and $p_s$ as in the
figs. \ref{fig1a}, \ref{fig1b}, \ref{fig3a}, \ref{fig3b}, \ref{fig5a},
\ref{fig5b}, \ref{fig7a}, and \ref{fig7b}.}}
\label{fig9}
\end{figure}
Therefore, we have explicitly computed the 
correlation $G_{0,0}$ as a function of the vehicle density, $c$, for 
the BJH model with $V_{max} = 1$ through computer simulation
(fig.~\ref{fig9}).  
This correlation falls exponentially with decreasing density and can 
be neglected if $c$ is not larger than about $0.3$. This is consistent 
with all the results summarized in
figs. \ref{fig1a}, \ref{fig1b}, \ref{fig3a}, \ref{fig3b}, \ref{fig5a},
\ref{fig5b}, \ref{fig7a}, and \ref{fig7b}, where we have 
observed reasonably good agreement between the analytical results and 
computer simulations for $c = 0.10$ and $0.25$ but not for $c = 0.50$.

\section{Summary and conclusion} 

There are, at present, several different CA models, each providing 
description of a range of traffic phenomena for which it has been 
specifically developed. Here we have considered the BJH model 
\cite{Benjamin96} which incorporates in a collective way, through 
the stochastic parameter $p_s$, a specific realistic aspect of 
individual driving habits into the minimal CA model suggested in the 
pioneering work of Nagel and Schreckenberg~\cite{Nagel92}. 

In this paper we have investigated in detail the spatio-temporal 
organization of the traffic in the BJH model \cite{Benjamin96} by 
calculating the distributions of DH, TH, jam sizes and gaps between 
jams. For $V_{max} = 1$, we have computed the DH distribution 
using the COMF approach. We have derived analytical expressions for 
the distributions of jam-sizes and gaps between jams as well as that 
of the TH distribution in the BJH model with $V_{max} = 1$  by a 
combination of COMF theory and a 2-cluster approximation within the 
SOMF theory. We have compared these approximate results with high 
quality numerical data obtained from our computer simulations. 
For $V_{max} > 1$, we have computed all the above mentioned 
distributions only via extensive computer simulations. 

We have analyzed the data for two different regimes, namely, 
$p_s \ll p$ and $p_s \gg p$. In the regime $p_s \ll p$, there is hardly 
any significant difference between the analytical results and 
computer simulation data for the BJH model. In the regime $p_s \gg p$, 
we have found, in general, good agreement between analytical results 
and computer simulation data for $V_{max} = 1$ when the density of 
vehicles is not too high. However, the higher is the density the wider 
is the difference between our analytical results and computer simulation 
data. This is a consequence of the fact that the COMF approximation 
as well as the 2-cluster SOMF  approximation made for our analytical 
calculations neglect some correlation which become non-negligible 
beyond a certain range of vehicle density. In order to directly 
establish this fact we have made a quantitative estimation of the 
regime of validity of our analytical results by computing an 
appropriate correlation function. 

In order to elucidate the effects of the slow-to-start rule on the 
spatio-temporal organization of the traffic, for each set of values 
of the parameters $V_{max}, p, c$, we have compared all our numerical 
data for the BJH model ($p_s \neq 0$) also with the corresponding data 
for the NaSch model ($p_s = 0$). From this comparison, some novel 
features of the spatio-temporal organization of vehicles in the BJH 
model have emerged. 

In the BJH model for $V_{max}=1$ no phase separation takes place and  
spontaneous jamming is possible also in the free-flow regime. In this 
sense the model is similar to the NaSch model. This is also reflected 
in the quantities we have calculated. For $V_{max} = 5$, at sufficiently 
low densities, i.e., in the free-flow regime, the behavior of the BJH 
model is also very similar to that of the NaSch model in the corresponding 
regime. But, at higher densities, phase-separated states appear in the 
BJH model with $V_{max} = 5$. Therefore, in this case, the jam-gap 
distribution is qualitatively different from that of the NaSch model, 
e.g., one finds a local maximum at a macroscopic distance corresponding 
to the length of the free-flow regime. On the other hand, the jam-size 
distribution is not a good indicator of the appearance of this phase 
separation because the large "jam", which is formed in this phase-separated 
regime contains 'holes' whereas, according to our definition, only 
a compact cluster of vehicles with $V = 0$ qualifies as a single jam. 

Measurements on real traffic have shown the existence of metastable
states~\cite{Kerner,Rehborn}. These are related to the phase separation
\cite{Barlovic98} found here and are an indication for the importance
of slow-to-start effects in real traffic. Slow-to-start rules lead
to a reduction of the outflow $J_{out}$ from a jam compared to the
maximal possible flow $J_{max}$ \cite{Barlovic98}. Experimentally
one finds $J_{max}/J_{out}\approx 1.5$ \cite{Kerner,Rehborn}. In the
BJH model for fixed $V_{max}$ this ratio is determined by $p$ and $p_s$.
$p_s$ also determines the velocity of jam fronts. Therefore it
is possible to tune these parameters in order to get a realistic
description of jam properties.

\vspace{1cm}

\noindent{\bf Acknowledgments:} It is our pleasure to thank D. Stauffer 
for a critical reading of the manuscript and for useful suggestions. 
One of the authors (DC) would like to thank International Center for 
Theoretical Physics, Trieste, for hospitality, where a part of the work 
was done. This work is supported, in part, by the SFB341 
K\"oln-Aachen-J\"ulich.

\newpage
\appendix
\setcounter{equation}{0}
\section{Details of the hybrid approach} 
\renewcommand{\theequation}{\Alph{section}.\arabic{equation}}

We now calculate the 2-cluster-like probabilities $p_2(a,b)$ involved 
in the hybrid approach to the BJH model with $V_{max} = 1$ immediately 
after the randomization stage from which one can obtain the conditional 
probabilities $p_2(\underbar{a}|b)$ etc.

\noindent$\bullet${\bf calculation of $p_2(-1,-1)$}: 
 
If the configuration of a pair of nearest-neighbor sites is $(-1,-1)$ after 
the acceleration stage, it does not change during the slow-to-start stage, 
blockage stage and the randomization stage and this observation leads to 
the relation  $p_2(-1,-1) = P_2(-1,-1)$. Using the 2-cluster-like 
decomposition, we write 
$$
{\cal P}(n) = P_2(\underline{+1}|-1) [P_2(\underline{-1}|-1)]^{n-1} 
P_2(\underline{-1}|1)
$$ 
for arbitrary $n \geq 2$ and, correspondingly, 
$$
{\cal P}(n+1) = P_2(\underline{+1}|-1) [P_2(\underline{-1}|-1)]^{n} 
P_2(\underline{-1}|1)
$$ 
which lead to 
${\cal P}(n+1)/{\cal P}(n) = P_2(\underline{-1}|-1) = 
P_2(-1,-1)/[P_2(-1,1)+P_2(-1,0)+P_2(-1,-1)]$. Using the result 
$P_2(-1,1) + P_2(-1,0) + P_2(-1,-1) = 1-c$ (because the probability that the 
left site is unoccupied, irrespective of the state of the right site, 
is $1-c$) 
we get ${\cal P}(n+1)/{\cal P}(n) = P_2(-1,-1)/(1-c)$. On the other hand, from 
the equation (\ref{eq3}) we get ${\cal P}(n+1)/{\cal P}(n) = z$. Therefore, 
\begin{equation}
p_2(-1,-1) = P_2(-1,-1) = z (1-c)
\label{eq41} 
\end{equation}

\noindent$\bullet${\bf calculation of $p_2(-1,+1)$}: 

If there there is a vehicle at the leading site (on the right) with speed $1$ 
at the end of the randomization stage then it must move forward during 
the next 
vehicle movement stage. Moreover, the vacancy on the left (following) site of 
the pair can be a part of s string of $m$ successive empty sites where $m$ can 
be any non-zero positive integer. Furthermore, since the DH distribution 
immediately after the randomization stage is identical to that immediately 
after the acceleration stage (as no vehicle movement takes place in between 
these two stages of the same time step of updating)  
\begin{equation}
p_2(-1,1) = c g \sum_{m=1}^{\infty}[{\cal P}(m) + {\tilde{\cal P}}(1)] 
= cg[1-{\cal P}(0)]
\label{eq42} 
\end{equation}
where we have utilized the normalization condition (\ref{eq1}). 

\noindent$\bullet${\bf calculation of $p_2(-1,0)$}: 

The calculation of $p_2(-1,0)$ proceeds exactly in the same manner as in case 
of $p_2(-1,+1)$ except for the fact that the leading site (on the right) 
is now occupied by a vehicle with speed $0$ which, consequently, will 
be unable to move
during the next vehicle movement stage; hence, the expression for $p_2(-1,0)$ 
is obtained from that for $p_2(-1,+1)$ replacing $g$ in $p_2(-1,+1)$ by $1-g$, 
i.e., 
\begin{equation}
p_2(-1,0) = c(1-g)[1-{\cal P}(0)]. 
\label{eq43} 
\end{equation}

\noindent$\bullet${\bf calculation of $p_2(0,-1)$}: 

The configuration $0,-1$ can arise immediately at the end of the randomization 
stage in three different ways:\\
(i) the vehicle in the following site (on the left) is one which is a 
slow-to-start vehicle which, indeed, slows down in the slow-to-start stage; 
the corresponding contribution to the probability $p_2(0,-1)$ is 
$c p_s {\tilde{\cal P}}(1)$.
(ii) the vehicle in the following site (on the left) is one which is a 
slow-to-start vehicle which, however does not slow down during the 
slow-to-start stage but slows down at the later stage of randomization; the 
corresponding contribution to $p_2(0,-1)$ is $c q_s p {\tilde{\cal P}}(1)$.
In both the cases (i) and (ii) the site immediately in front of the leading 
site (on the right) must be occupied in order that the vehicle at the 
following 
site of the pair can qualify as a candidate for slow-to-start option.
(iii) the vehicle in the following site is not a slow-to-start vehicle and it 
slows down at the randomization stage. Moreover, in this case, the empty 
leading site can be a part of string of $m$ successive empty sites where $m$ 
can be any non-zero positive integer; the corresponding contribution to 
$p_2(0,-1)$ is 
$c p [\sum_{m=1}^{\infty} {\cal P}(m)] = c p [1 - {\cal P}(0) - {\tilde 
{\cal P}}(1)]$. 
Therefore, adding all the three contributions we get 
\begin{equation}
p_2(0,-1) = c [p_s {\tilde{\cal P}}(1) + q_s p {\tilde {\cal P}}(1) 
+ p (1 - {\cal P}(0) - {\tilde {\cal P}}(1))].
\label{eq44} 
\end{equation}

\noindent$\bullet${\bf calculation of $p_2(0,0)$}: 

The configuration $0,0$ can be obtained immediately after the randomization 
stage if and only if it was $+1,+1$ immediately after the acceleration stage. 
The facts (i) that the following site is occupied by a vehicle, (ii) that 
there is no empty site in front of the following vehicle, and (iii) that the 
leading vehicle cannot move during the vehicle movement stage (because of its 
speed being $0$ at that stage) lead to 
\begin{equation}
p_2(0,0) = c (1-g) {\cal P}(0) 
\end{equation}

\noindent$\bullet${\bf calculation of $p_2(0,+1)$}: 

We now apply the same arguments as those in the calculation of $p_2(0,0)$ 
except that now the leading vehicle will move during the next vehicle 
movement stage; therefore, replacing $1-g$ in the expression for $p_2(0,+1)$ 
by $g$ we get 
\begin{equation}
p_2(0,+1) = c g {\cal P}(0).
\label{eq46} 
\end{equation} 

\noindent$\bullet${\bf calculation of $p_2(1,-1)$}: 

The configuration $+1,-1$ can arise immediately at the end of the 
randomization stage in two different ways:\\
(i) the vehicle in the following site (on the left) is one which is a 
slow-to-start vehicle which, however neither slows down during the 
slow-to-start stage nor slows down at the later stage of randomization; the 
corresponding contribution to $p_2(+1,-1)$ is $c q_s q {\tilde{\cal P}}(1)$.
In this case (i) the site immediately in front of the leading site 
(on the right) must be occupied in order that the vehicle at the following 
site of the pair can qualify as a candidate for slow-to-start option.
(ii) the vehicle in the following site is neither a slow-to-start vehicle and  
nor does it slow down at the randomization stage. Moreover, in this case, the 
empty leading site can be a part of string of $m$ successive empty sites where 
$m$ can be any non-zero positive integer; the corresponding contribution to 
$p_2(+1,-1)$ is 
$c q \sum_{m=1}^{\infty} {\cal P}(m) = c q [1 - {\cal P}(0) - 
{\tilde {\cal P}}(1)]$. 
Therefore, adding the two contributions we get 
\begin{equation}
p_2(+1,-1) = c q [q_s {\tilde {\cal P}}(1) + (1 - {\cal P}(0) - 
{\tilde {\cal P}}(1))].
\label{eq47} 
\end{equation}

\noindent$\bullet${\bf calculation of $p_2(+1,0)$ and $p_2(+1,+1)$}: 
Since the speed of a following vehicle is certainly reduced to zero at the 
blocking stage if the leading site is occupied by a vehicle, it is impossible 
to have the configurations $+1,0$ and $+1,+1$ immediately after the 
randomization stage; therefore, 
\begin{equation}
p_2(+1,0)= 0 
\label{eq48} 
\end{equation}
and 
\begin{equation}
p_2(+1,+1) = 0. 
\label{eq49} 
\end{equation}  
In the special case $p_s = 0$ the 2-cluster-like probabilities 
(\ref {eq41})-(\ref{eq49}) reduce to the corresponding expressions 
for the true 2-cluster probabilities derived earlier
\cite{Chowdhury97} for the  NaSch model.

\setcounter{equation}{0}
\section{The 2-cluster equations} 

Using the symbols $\mu = p_s + p q_s$ and $\nu = q q_s$ we can write the 
exact equations for time evolution of the 2-cluster probabilities 
$C_2(s_1,s_2)$ in the BJH model in terms of the 2-cluster and 
higher-cluster probabilities in the steady state as follows. 
\begin{eqnarray}
C_2(-1,-1) &=& C_3(-1,-1,-1) + pC_3(0,-1,-1) + pC_3(1,-1,-1) 
                 + \mu C_3(2,-1,-1) \nonumber \\
  &+& qC_4(-1,-1,0,-1) + pqC_4(0,-1,0,-1) + \mu qC_4(2,-1,0,-1) \nonumber\\
  &+& pqC_4(1,-1,0,-1) + \nu C_4(-1,-1,2,-1) + p\nu C_4(0,-1,2,-1)\nonumber\\
  &+& \mu\nu C_4(2,-1,2,-1) + p\nu C_4(1,-1,2,-1) + qC_4(-1,-1,1,-1)\nonumber\\
  &+& pqC_4(0,-1,1,-1) + \mu qC_4(2,-1,1,-1) + pqC_4(1,-1,1,-1)\\
\label{eq50}
C_2(0,-1) &=&pC_2(0,-1)+\mu C_2(2,-1)+pC_2(1,-1)\\ 
\label{eq51}
C_2(2,-1) &=& qC_2(0,-1) + \nu C_2(2,-1) - qC_3(-1,0,-1) - 
                 \nu C_3(-1,2,-1)\nonumber \\
 &+& qC_3(0,0,-1) + \nu C_3(0,2,-1) + qC_3(0,1,-1) + qC_3(2,1,-1)\nonumber\\
 &+& qC_3(1,1,-1) - qC_3(0,0,-1) - \nu C_3(0,2,-1)\\
\label{eq52}
C_2(1,-1) &=& qC_3(0,-1,-1)+qC_3(1,-1,-1)+\nu C_3(2,-1,-1) 
      + q^2C_4(0,-1,0,-1) \nonumber \\
 &+& q\nu C_4(2,-1,0,-1) + q^2C_4(1,-1,0,-1)+q\nu C_4(0,-1,2,-1) \nonumber\\
 &+& \nu^2 C_4(2,-1,2,-1)+q\nu C_4(1,-1,2,-1) + q^2C_4(0,-1,1,-1)\nonumber\\
 &+& q\nu C_4(2,-1,1,-1)+q^2C_4(1,-1,1,-1)\\
\label{eq53}
C_2(-1,0) &=& pC_4(-1,-1,0,-1) + p^2C_4(0,-1,0,-1) 
                + \mu pC_4(2,-1,0,-1) \nonumber \\
 &+& p^2C_4(1,-1,0,-1) + \mu C_4(-1,-1,2,-1) + p\mu C_4(0,-1,2,-1)\nonumber\\
 &+& \mu^2 C_4(2,-1,2,-1) + p\mu C_4(1,-1,2,-1) + pC_4(-1,-1,1,-1)\nonumber\\ 
 &+& p^2C_4(0,-1,1,-1) + p\mu C_4(2,-1,1,-1) + p^2C_4(1,-1,1,-1)
\label{eq54}
\end{eqnarray}
\begin{eqnarray}
 C_2(0,0)&=&0\\
\label{eq55}
 C_2(2,0) &=& pC_2(0,-1)+\mu C_2(2,-1)+pC_2(1,-1)-pC_3(-1,0,-1) \nonumber\\
      &-& \mu C_3(-1,2,-1)-pC_3(-1,1,-1)\\
\label{eq56}
C_2(1,0) &=& pqC_4(0,-1,0,-1)+p\nu C_4(2,-1,0,-1) 
             + pqC_4(1,-1,0,-1) \nonumber \\ 
 &+& q\mu C_4(0,-1,2,-1) + \mu \nu C_4(2,-1,2,-1) + \mu qC_4(1,-1,2,-1) 
       \nonumber \\
 &+& pqC_4(0,-1,1,-1)+p\nu C_4(2,-1,1,-1) + pqC_4(1,-1,1,-1)\\
\label{eq57}
C_2(-1,2) &=& C_3(-1,-1,0) - C_4(-1,-1,0,-1) + pC_3(0,-1,0) 
            - pC_4(0,-1,0,-1) \nonumber \\ 
 &+& \mu C_3(2,-1,0)-\mu C_4(2,-1,0,-1) + pC_3(1,-1,0) - pC_4(1,-1,0,-1) 
      \nonumber \\ 
 &+& C_3(-1,-1,2) - C_4(-1,-1,2,-1) + pC_3(0,-1,2) - pC_4(0,-1,2,-1) 
      \nonumber \\
 &+& \mu C_3(2,-1,2) - \mu C_4(2,-1,2-1) + pC_3(1,-1,2) - pC_4(1,-1,2,-1) 
      \nonumber \\ 
 &+& C_3(-1,-1,1)-C_4(-1,-1,1,-1) + pC_3(0,-1,1) - pC_4(0,-1,1,-1) 
       \nonumber \\ 
 &+& \mu C_3(2,-1,1) - \mu C_4(2,-1,1,-1)+pC_3(1,-1,1)-pC_4(1,-1,1,-1)
\nonumber\\
\label{eq58}
\end{eqnarray}
\begin{eqnarray}
 C_2(0,2)&=&0\\ 
\label{eq59} 
C_2(2,2) &=& C_2(2,0) + C_2(1,0) + C_2(2,2) + C_2(1,2) - C_3(2,0,-1) 
         \nonumber \\
 &-& C_3(1,0,-1) - C_3(2,2,-1) - C_3(1,2,-1) + C_2(0,0)  \nonumber \\ 
 &-& C_3(0,0,-1) + C_2(0,2) - C_3(0,2,-1) + C_2(0,1) -C_3(0,1,-1)\nonumber \\ 
 &+& C_2(2,1)-C_3(2,1,-1) + C_2(1,1) - C_3(1,1,-1)\\
\label{eq60}
C_2(1,2) &=& qC_3(0,-1,0) - qC_4(0,-1,0,-1) + \nu C_3(2,-1,0) 
          - \nu C_4(2,-1,0,-1) \nonumber \\ 
 &+& qC_3(1,-1,0) - qC_4(1,-1,0,-1) + qC_3(0,-1,2) - qC_4(1,-1,2,-1)
            \nonumber\\ 
 &+& \nu C_3(2,-1,2) - \nu C_4(2,-1,2,-1) + qC_3(1,-1,2) - qC_4(1,-1,2,-1) 
             \nonumber \\
 &+& qC_3(0,-1,1) - qC_4(0,-1,1,-1) + \nu C_3(2,-1,1) - \nu C_4(2,-1,1,-1) 
              \nonumber \\ 
 &+& qC_3(1,-1,1)-qC_4(1,-1,1,-1)\\
\label{eq61}
C_2(-1,1) &=& qC_2(0,-1) + \nu C_2(2,-1) + qC_2(1,-1)\\
\label{eq62}
C_2(0,1)&=&0\\
\label{eq63}
C_2(2,1)&=&0\\
\label{eq64}
C_2(1,1)&=&0.
\label{eq65}
\end{eqnarray}
In order to close these equations, we now make the approximate 
decomposition of the 3-cluster and 4-cluster probabilities in 
terms of 2-cluster probabilties using the formulae
\begin{equation}
C_3(x_1,x_2,x_3)=\frac{C_2(x_1,x_2)C_2(x_2,x_3)}{\sum_{s=-1}^2 C_2(x_2,s)}
\label{eq66}
\end{equation}
\begin{equation}
C_4(x_1,x_2,x_3,x_4)=\frac{C_2(x_1,x_2)C_2(x_2,x_3)C_2(x_3,x_4)}{\left[
\sum_{s=-1}^2 C_2(s,x_2)\right]\cdot\left[\sum_{s=-1}^2 C_2(x_3,s)\right]} 
\label{eq67}
\end{equation}
We like to remark that the so-called Kolmogorov consistency
conditions \cite{Gutowitz87}
\begin{equation}
\sum_{s=-1}^2 C_2(-1,s)=\sum_{s=-1}^2 C_2(s,-1)=1-c
\end{equation}
and
\begin{equation}
\sum_{s=-1}^2 \left[C_2(0,s)+C_2(1,s)+C_2(2,s)\right]
=\sum_{s=-1}^2 \left[C_2(s,0)+C_2(s,1)+C_2(s,2)\right]=c
\end{equation}
are useful also for the numerical solution of the equations 
(\ref{eq50}-\ref{eq65}).
\setcounter{equation}{0}
\section{Relation between COMF and hybrid approaches}

In order to point out the equivalence of the hybrid  and COMF 
approaches, we sketch here the derivation of the distribution of 
the jam sizes in the BJH model with $V_{max} = 1$ following the 
hybrid approach and show that we get the same expression as we 
derived in section 4 following the pure COMF approach. 

Decomposing the distribution ${\cal P}_{js}(k)$ into the 2-cluster-like 
conditional probabilities $p_2(\underbar{a}|b)$ we get 
\begin{equation}
{\cal P}_{js}(k) = {\cal N}_{js} p_2(\underbar{-1}|0)
[p_2(\underbar{0}|0)]^{k-1} p_2(\underbar{0}|-1) + p_2(\underbar{-1}|0)
[p_2(\underbar{0}|0)]^{k-1} p_2(\underbar{0}|1) 
\label{eq68}
\end{equation}
where ${\cal N}_{js}$ is a normalization coefficient. Using the 
expressions for the 2-cluster-like probabilities derived in the 
appendix A and normalizing the distribution to unity we recover 
\cite{Sinha98} the expression (\ref{eq12}) of ${\cal P}_{js}(k)$ 
in terms of ${\cal P}(0)$.

\end{document}